\documentclass[preprint, superscriptaddress, showpacs]{revtex4}

\usepackage[dvips]{graphicx}
\usepackage{color}
\usepackage{amssymb}
\usepackage{amsmath}
\usepackage[latin1]{inputenc} % to allow German characters as {\"a} to be typeset correctly

\newcommand{\ep}{\varepsilon}
\newcommand{\s}{\mathtt{s}}
\newcommand{\p}{\mathtt{p}}
\newcommand{\rss}{r_{\s\s}}
\newcommand{\rsp}{r_{\s\p}}
\newcommand{\rps}{r_{\p\s}}
\newcommand{\rpp}{r_{\p\p}}

\newcommand{\LambdaMOKE}{\Lambda_\mathrm{MOKE}}
\newcommand{\lambdaperp}{\lambda_{\mathrm{sw},\perp}}

\newcommand{\mLnull}{m_L^{(0)}}
\newcommand{\mPnull}{m_P^{(0)}}
\newcommand{\kswz}{k_{z,\mathrm{sw}}}

\newcommand{\omsw}{\omega_\mathrm{sw}} %sw frequency
\newcommand{\oml}{\omega_0}  % light frequency

\def\baf{\discretionary{}{}{}} % gives possible breaks in the word, but the break of the word is provided withoud adding the dash

\begin{document}

\title{Analytical expression of the magneto-optical Kerr effect and Brillouin light scattering intensity arising from dynamic magnetization}

\author{Jaroslav Hamrle}
\email[Corresponding author:~]{jaroslav.hamrle@vsb.cz}
\affiliation{Centre for Advanced Innovation Technology, VSB - Technical University of Ostrava, 708 33 Ostrava-Poruba, Czech Republic}
\affiliation{Fachbereich Physik and Forschungszentrum OPTIMAS,
Technische Universit\"at Kaiserslautern,
Erwin-Schr\"odinger-Stra\ss e 56, D-67663 Kaiserslautern, Germany}
\author{Jarom\'\i r Pi\v{s}tora}
\affiliation{Department of Physics, VSB - Technical University of Ostrava,
708 33 Ostrava-Poruba, Czech Republic}
\author{Burkard Hillebrands}
\affiliation{Fachbereich Physik and Forschungszentrum OPTIMAS,
Technische Universit\"at Kaiserslautern,
Erwin-Schr\"odinger-Stra\ss e 56, D-67663 Kaiserslautern, Germany}

\author{Benjamin~Lenk}
\author{Markus~M\"{u}nzenberg}
\affiliation{Institute of Physics, University of G\"{o}ttingen, 37077~G\"{o}ttingen, Germany}

\begin{abstract}

Time-resolved magneto-optical Kerr effect (MOKE) and Brillouin light scattering (BLS) spectroscopy are important techniques for the investigation of magnetization dynamics. Within this article, we calculate analytically the MOKE and BLS signals from prototypical spin-wave modes in the ferromagnetic layer. The reliability of the analytical expressions is confirmed by optically exact numerical calculations. Finally, we discuss the dependence of the MOKE and BLS signals on the ferromagnetic layer thickness.

\end{abstract}

% PACS
% 75.30.Ds: Magnetic properties: spinwaves
% 75.50.Cc: ferromagnetic materials: metals other than iron
% 78.35.+c: Light scattering: in condensed metter
% 78.20.Ls: Magneto-optical effects

\maketitle

\section{Introduction}

Optical techniques based on the magneto-optical Kerr effect (MOKE), such as time-resolved MOKE (TR-MOKE) or Brillouin light scattering (BLS) spectroscopy are routinely used for the investigation of magnetization dynamics. Whereas TR-MOKE provides information on the magnetization dynamics within the time domain, BLS informs on magnetization dynamics within the frequency domain.

BLS measures light intensity scattered by a spin wave. In a~phenomenological (wave) picture a~spin wave is a periodical displacement of the magnetization with respect to the saturated state, oscillating at a~spin-wave frequency $\omsw$, and propagating at a given spin-wave $k$-vector.
The spin waves result from the coupling between the spins, dominated by exchange and dipolar interactions \cite{kal86, sta91,hei93}.
Any periodical variation of the optical properties (namely, periodic variation of the permittivity tensor $\varepsilon(t,\vec{r})$) works as an effective oscillating and propagating optical grating \cite{coch88, coch90, gio01, buch07}. Hence, due to the spatial periodicity of the effective grating, the reflected light is scattered and changes its propagation direction. Furthermore, the oscillation of the effective grating changes the light frequency of the scattered light, a quantity detected in a BLS spectrometer \cite{moc87}. In a (pseudo-)particle picture, the scattering of the light by the spin wave can be interpreted as inelastic scattering of photons on magnons \cite{fle67}, where the photon is gaining (loosing) its energy as it absorbes (creates) a magnon, respectively. Therefore, in contrast to TR-MOKE, BLS can also detect non-coherent spin waves such as thermal spin waves. Note, that the BLS technique usually provides a larger experimental sensitivity when compared to the TR-MOKE technique. For example, it has been demonstrated to detect thermal spin waves on a Co monolayer \cite{kra92}.

TR-MOKE investigations are based on repeated excitations of magnetization dynamics, usually by magnetic field pulses or by intense light pulses \cite{fre02, kim07, neu08}. The resulting excitations are then stroboscopically detected making use of the MOKE. Therefore, TR-MOKE provides an insight into the magnetization dynamics  within the time domain. Using Fourier transformation, the TR-MOKE signal can be easily transformed to the frequency domain \cite{bue04}. Hence, in case of externally excited systems, TR-MOKE and BLS investigations are complementary. This was nicely demonstrated by K.\ Perzlmaier \textit{et al} investigating confined spin-wave modes in a square permalloy element \cite{per05}.

The numerical models to calculate the BLS light intensity were elaborated by J.R.\ Cochran \textit{et al} \cite{coch88, coch90}, followed by L.\ Giovannini \textit{et al} \cite{gio01}. Later, a~simple relation between the complex Kerr angle and the BLS intensity was expressed by M.\ Buchmeier \textit{et al} \cite{buch07}. However, all those treatments of the BLS intensity are numerical ones, and up to now, there exists no analytical expression of the MOKE and BLS signals originating from spin-wave modes.

Within this article, we present the analytical dependence of the TR-MOKE and BLS signals for several types of spin-wave modes. Such calculations can serve either for separation of MOKE and BLS signals from a single ferromagnetic (FM) material in a stack of FM layers \cite{fer97, ham02, pos08, pos09}, or for the quantitative determination of the energy carried by each spin-wave mode.

The article is organized as follows: In Sec.~\ref{Sec:II} we establish a relation between the BLS intensity and strength of the MOKE effect. In Sec.~\ref{Sec:III} the analytical expression of the MOKE depth sensitivity function is developed, and we discuss its validity and properties. Section~\ref{Sec:IV} provides the analytical expressions of the MOKE and BLS signals. Finally Section~\ref{Sec:V} compares analytical expressions with optically exact numerical models, and we discuss in detail the FM-layer thickness dependence of the MOKE and BLS signals.

\section{Relation between MOKE and BLS signals}
\label{Sec:II}

As the light travels through the FM layer, the light intensity is attenuated and the phase of the light is delayed. Therefore, sublayers of the FM material situated at different depths of the FM layer provide a different MOKE response to a given magnetization state. Due to the fact that the MOKE is linear in magnetization, the result can be written as a superposition of the single contributions coming from the different depths of the FM layer \cite{zak90,hub93,kam98,ham02}:
\begin{equation}
\label{eq:Phimode}
\Phi_{\s/\p}(t)=\int_0^d
\left[
L_{\s/\p}(z) m_L(z,t) + P_{\s/\p}(z) m_P(z,t)
\right] \mathrm{d}z=\Phi_{\s/\p} \sin(\omsw t +\phi_{\s/\p}),
\end{equation}
where $\Phi_\s=\rps/\rss$ and $\Phi_\p=-\rsp/\rpp$ are the $\s$- and $\p$- complex Kerr angles, arising from the FM film when the incident light is $\s$ and $\p$ polarized, respectively. The terms $r_{xy}$, $x,y=\{\s,\p\}$ stand for components of the reflection matrix. $L_{\s/\p}(z)$ and $P_{\s/\p}(z)$ are the complex MOKE depth sensitivity functions related to longitudinal and polar magnetization, respectively,  $d$ being the FM layer thickness. $m_L(z,t)$ and $m_P(z,t)$ are the depth profiles of the magnetizations in the FM film having longitudinal (i.e.\ in-plane and parallel to the plane of the light incidence) and polar (i.e.\ normal) directions. In the case of spin waves, those magnetizations correspond to profiles of the dynamic magnetization (i.e.\ spin-wave amplitudes), precessing at the frequency $\omsw$,
\begin{align}
\label{eq:mlmp}
m_P(z,t)&=m_P(z)\sin\omsw t \\
m_L(z,t)&=m_L(z)\cos\omsw t.
\end{align}

According to Refs.~\cite{buch07,ham09cms}, the BLS intensity $I^{(\oml\pm\omsw)}_{\s/\p}$ of the backscattered light can be expressed in a rather similar way as the complex Kerr angle $\Phi$:
 \begin{equation}
\label{eq:BLS}
I^{(\oml\pm\omsw)}_{\s/\p}=
I_0 \left|
\int_0^d r_{\s\s/\p\p} \left[-L_{\s/\p}(z) m_L(z) + P_{\s/\p}(z)m_P(z)
\right] \mathrm{d}z
\right|^2.
\end{equation}
Within the depth sensitivity, the main difference between the complex Kerr angle $\Phi$ and the BLS intensity $I^{(\oml\pm\omsw)}_{\s/\p}$ is given by the fact that whereas MOKE is a linear combination of $L_{\s/\p}$, $P_{\s/\p}$ and the magnetization profiles $m_L(z)$, $m_P(z)$, respectively, the BLS intensity has a quadratic form.

From comparing the equations expressing the complex Kerr angle (Eq.~\ref{eq:Phimode}) and the BLS intensity (Eq.~\ref{eq:BLS}), their close similarity is apparent. With exception of the sign of the longitudinal contribution, the BLS intensity is basically a~quadratic form of MOKE. Therefore, the BLS intensity can be  expressed as being proportional to the square of the off-diagonal reflection coefficients $r_{\s\p/\p\s}$
\begin{equation}
\label{eq:MOKEBLS}
I_{\s/\p}^{(\oml\pm\omsw)}
= I_0\left|
r_{\p\s/\s\p}
\right|^2
\equiv I_0 \left| \pm
r_{\s\s/\p\p} \Phi_{\s/\p}
\right| ^2,
\end{equation}
where we must reverse the sign of the longitudinal contribution when expressing $\Phi_{\s/\p}$. For example, this can be achieved either by reversing the sign of  $m_L$ in the calculations.
The complex Kerr angle (i.e.\ the magnitude of the MOKE) is denoted by $\Phi_{\s/\p}$, neglecting its time dependence (i.e.\  omitting the term $\sin(\omsw t+\phi_{\s/\p})$ in Eq.~(\ref{eq:Phimode})).

%However, as follow from expressions of MOKE intensities (later derived (Eqs.~(\ref{eq:MOKEde}, \ref{eq:MOKEdebot}, \ref{eq:MOKEpssw}))), opposite sign of longitudinal contribution (i.e. change of sign of $\gamma$) changes sign only phase of the detected wave $\phi_{\s/\p}$ of the detect MOKE signal, keeping MOKE effects itself $\Phi_{\s/\p}^\mathrm{(DE1)}$, $\Phi_{\s/\p}^\mathrm{(DE2)}$, $\Phi_{\s/\p}^\mathrm{(PSSW)}$ unchanged .

\section{Analytical expression of the MOKE depth sensitivity function}
\label{Sec:III}

In the case of an optically thick FM layer (i.e.\ the FM film thickness $d$ is larger than the MOKE probing depth $\LambdaMOKE=\lambda/(4\pi \mathrm{Im}(N_z))$, $\lambda$ being the vacuum light wavelength and $N_z$ defined just below), $L_{\s/\p}(z)$ and $P_{\s/\p}(z)$ can be analytically expressed as \cite{hub93}
\begin{align}
\label{eq:dsP}
P_{\s/\p}(z)&=P_{\s/\p}(0) \exp[-4\pi i N_z z/\lambda]
\\
\label{eq:dsL}
L_{\s/\p}(z)&=L_{\s/\p}(0) \exp[-4\pi i N_z z/\lambda],
\end{align}
where we define $\gamma_{\s/\p}$ to be the ratio of the LMOKE and PMOKE response at the upper interface (i.e.\ $z=0$) of the FM layer, $L_{\s/\p}(0)=\gamma_{\s/\p} P_{\s/\p}(0)$,
$N_z$ is the normalized wave-vector of light in polar (i.e.\ normal) direction, $N_z=\sqrt{(N^\mathrm{(fm)})^2-(N^\mathrm{(air)}\sin\varphi)^2}$, where $N^\mathrm{(fm)}$ and $N^\mathrm{(air)}$ are the refractive indices of the FM layer and air, respectively, and $\varphi$ is the angle of light incidence with respect to the sample normal, respectively. In general, metals provide a relatively large value of the optical permeability $\epsilon_0\equiv N^2$, $N$ being the refractive index. Therefore, the angular dependence of $N_z$ can be neglected and hence  $N_z\approx N$. For example, for Ni at $\lambda=810$\,nm, $\ep^\mathrm{(Ni)}_0=-13.24 +22.07i$ and hence $N_z$ is reduced only by 1\% when going from $\varphi=0$ to 90$^\circ$ \cite{ham02}.

\begin{figure}
\begin{center}
\includegraphics[width=0.9\textwidth]{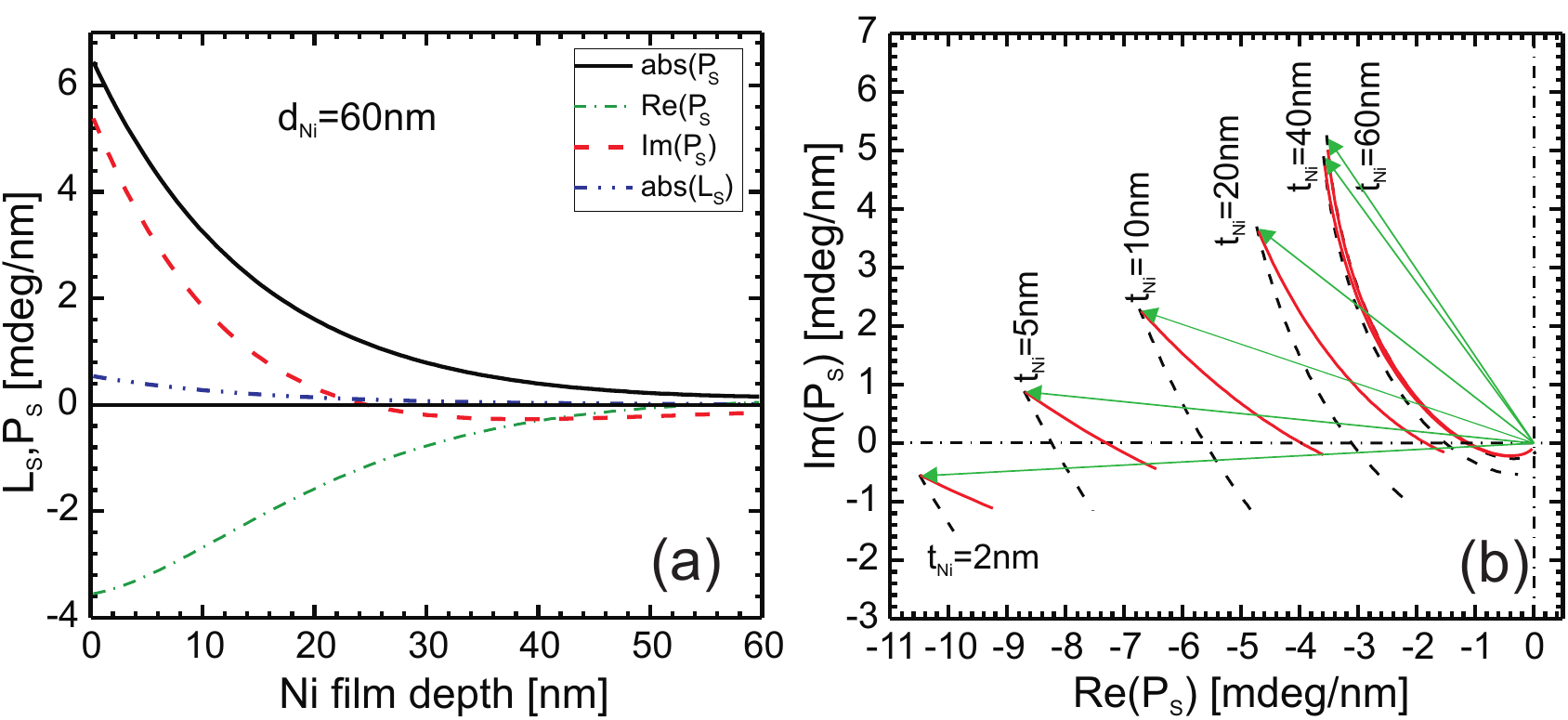}
\end{center}
\caption{%
\label{f:MOKE}
(color) (a) Depth dependence of the longitudinal and the polar depth sensitivity function $L_\s$, $P_\s$ inside the 60\,nm-thick Ni film. (b) Depth sensitivity function $P_s$ for Ni-thicknesses 10, 20, 40 and 60\,nm, presented in the complex plane calculated by the $4\times4$ matrix formalism (dashed black line) and using the analytical expression Eq.~(\ref{eq:dsP}) (solid red line). Light wavelength and incidence angle are $810$\,nm and 25$^\circ$, respectively. Green arrows point the start point of the depth sensitivity function, $P_{\s}(0)$, i.e.\ its value at $z=0$.}
\end{figure}

An example of $L_{\s/\p}(z)$ and $P_{\s/\p}(z)$ is shown in Fig.~\ref{f:MOKE}(a) for the multilayer structure air/Cu(2\,nm)/\baf Ni(60\,nm)/\baf Cu(5\,nm)/\baf Si. The calculations were done using a~$4\times 4$ matrix formalism \cite{vis91}, for a~light wavelength of $\lambda=810$\,nm and an incidence angle of $\varphi=25^\circ$. The diagonal permittivity of the used materials are
$\ep^\mathrm{(Cu)}_0=-26.37 + 2.61i $,
$\ep^\mathrm{(Ni)}_0=-13.24 +22.07i$,
$\ep^\mathrm{(Si)}_0=13.58 + 0.04i$, the off-diagonal permittivity of Ni being $\ep^\mathrm{(Ni)}_{\mathrm{off}}=0.217-0.091i$ \cite{palik}. As the difference between $L_\s$ and $L_\p$ ($P_\s$ and $P_\p$) is only the starting amplitude and phase of $L_{\s/\p}(0)$ and $P_{\s/\p}(0)$, the calculations in Fig.~\ref{f:MOKE}(a) are presented only for $L_\s$ and $P_\s$. In this particular case, the polar MOKE (PMOKE) amplitude is about 12$\times$ larger than the longitudinal MOKE (LMOKE) amplitude. It can be considered a general rule that the PMOKE is stronger than the LMOKE. Additionally, in our example, also the small incidence angle of $\varphi=25^\circ$ contributes to the small value of the LMOKE (LMOKE vanishes at $\varphi=0$). But even for an incidence angle of about 60-70$^\circ$, when the LMOKE reaches its maximum, its amplitude would increase only by a factor of 2, still much smaller than the PMOKE.

Figure~\ref{f:MOKE}(b) shows an agreement between the analytical expressions of the MOKE depth sensitivity functions $L_{\s/\p}(z)$, $P_{\s/\p}(z)$ as given by Eqs.~(\ref{eq:dsP}--\ref{eq:dsL}) and as determined from optically exact $4\times4$ matrix calculations for various Ni thicknesses. The starting point, $L_{\s/\p}(0)$, $P_{\s/\p}(0)$ is determined by the optical $4\times$4 calculations in both cases.
It is demonstrated that there is nearly a~perfect agreement for large Ni thicknesses 40 and 60\,nm, whereas there is a disagreement for Ni thicknesses below 30\,nm. It is because the analytical expressions of $L_{\s/\p}(z)$ and $P_{\s/\p}(z)$ are valid when the thickness of the FM layer $d$ is larger than the MOKE probing length, $\LambdaMOKE=\lambda/(4\pi \mathrm{Im}(N_z))$ ($\LambdaMOKE=14.5$\,nm for Ni at $\lambda=810$\,nm). On the other hand, for very small thicknesses of the FM layer (when $d\ll\LambdaMOKE$, i.e.\ below $d=3$\,nm), the FM layer can be neglected from the optical point of view. Hence,  $P_{\s/\p}$, $L_{\s/\p}$ can be considered constant so that the expressions of $L_{\s/\p}$, $P_{\s/\p}$ are also valid for a very small thicknesses of the FM layer (so-called ultrathin FM layer approximation \cite{vis95}).

Therefore, the MOKE depth sensitivity functions $P_{\s/\p}$, $L_{\s/\p}$ are described well by the analytical expressions (Eqs.~(\ref{eq:dsP}--\ref{eq:dsL})) with exception of the FM thickness range between about 3\,nm to 30\,nm. However, even in this range, the analytical expression describes basic features of the depth sensitivity functions. Namely a reduction of $|P_{\s/\p}|$, $|L_{\s/\p}|$ and an increase of the phase $\arg(P_{\s/\p})$, $\arg(L_{\s/\p})$ with increasing depth inside FM film. Hence, deviation of the simple analytical calculations from an optically exact calculations is not large even in case of this thickness interval, as shown later.

We finally note that $L_{\s/\p}(z)$, $P_{\s/\p}(z)$ are nearly independent of the incidence angle as their angular dependence is governed solely by $N_z$, whose angular dependence is very weak in the general case of metals.

% is much smaller than thickness of the FM layer (i.e.\ effectively, the FM layer is optically infinite). Note that penetration depth of light $\Lambda_\mathrm{pen}=\lambda/(2\pi\mathrm{Im}(N_z))=2\Lambda_\mathrm{MOKE}$, as in case of MOKE the light has to travel to a given sublayer and back.

%The analytical expressions of $L_{\s/\p}$, $P_{\s/\p}$ (Eqs.~(\ref{eq:dsP}-\ref{eq:dsL})) show that their depth dependence is described by an exponencial decay accompanied by a  linear phase shift. Hence, with increasing depth, $|P_{\s/\p}(z)|$ and $|L_{\s/\p}(z)|$ exponentially decay, whereas their real and imaginary parts (corresponding to Kerr rotation and Kerr ellipticity) can pass zero or change sign due to the increasing phase shift. This provides a possibility to cancel the MOKE or BLS signal from an arbitrary FM layer in case of a stack of the FM layers \cite{ham02}.

\section{MOKE and BLS from Damon-Eshbach and Perpendicular Standing Spin-Wave Modes}
\label{Sec:IV}

To obtain the MOKE or BLS response of a given spin-wave mode, the profile of the dynamic magnetizations $m_L(z)$, $m_P(z)$ through the FM film must be determined first. Those calculations are usually based on phenomenological models of the magnetization inside the FM layer \cite{kal86,hil90,buch03,buch07}. Then, the complex Kerr angle $\Phi$ or the BLS intensity $I^{(\oml\pm\omsw)}$ coming from a given spin-wave mode can be expressed using Eq.~(\ref{eq:Phimode}) and (\ref{eq:BLS}), respectively.

\begin{figure}
\begin{center}
\includegraphics[width=0.4\textwidth]{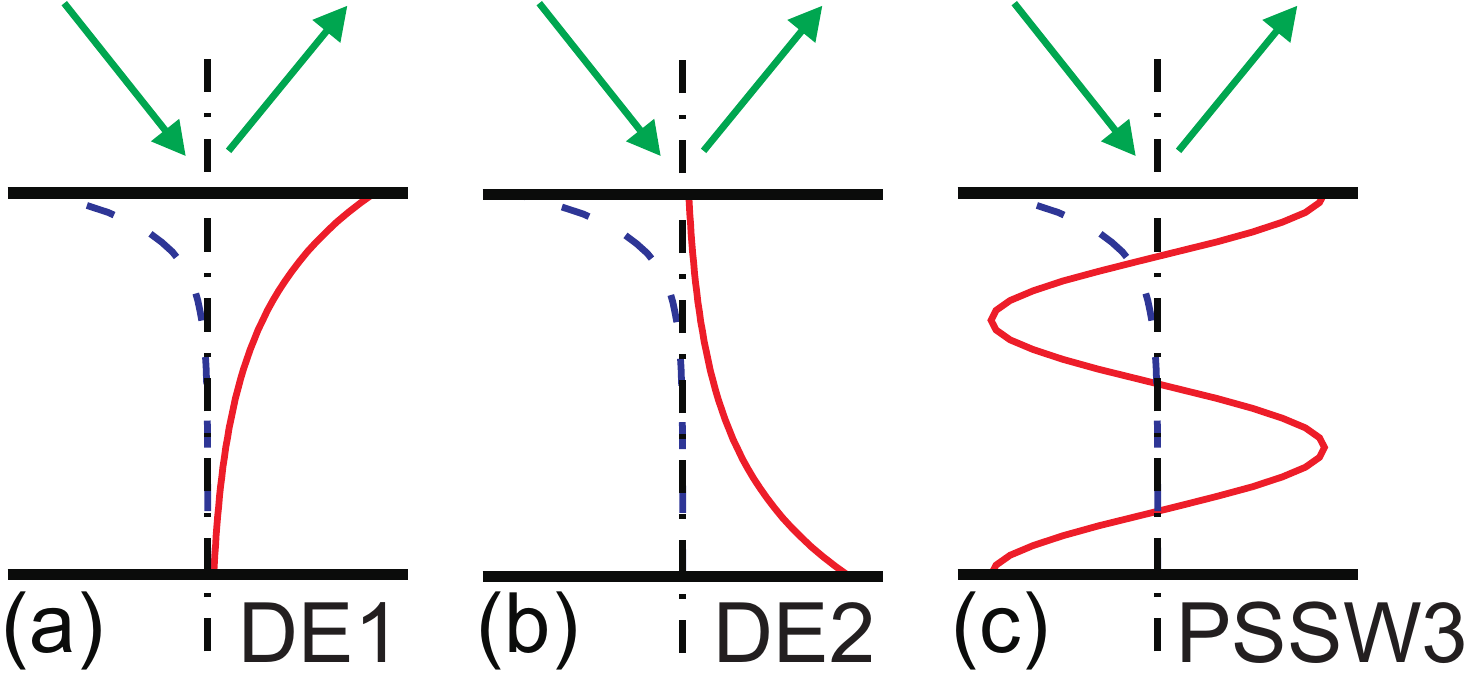}
\end{center}
\caption{%
\label{f:sketch}%
(color) (solid red line) The sketches of the treated spin-wave modes (a) DE mode bounded to the upper interface (called DE1) (b) DE mode bounded to the lower interface (called DE2) (c) PSSW mode. (blue dashed line) Sketch of the MOKE depth sensitivity function.
}
\end{figure}

In general, the profile of a~spin-wave mode must be calculated numerically. However, prototypical spin-waves modes have rather simple analytical expressions of their amplitude profiles. Here we work out the magneto-optical response of three spin-wave modes, the Damon-Eshbach (DE) mode (including the homogeneous ferromagnetic resonance (FMR) mode) bounded to the upper (Fig.~\ref{f:sketch}(a)) and lower (Fig.~\ref{f:sketch}(b)) interface, and the perpendicular standing spin-wave (PSSW) mode (Fig.~\ref{f:sketch}(c)).

The DE mode occurs when $\vec{M}$ lies in-plane and the spin wave propagates in  a direction perpendicular to $\vec{M}$. This mode can be bounded either to the upper (Fig.~\ref{f:sketch}(a)) or lower (Fig.~\ref{f:sketch}(b)) interface, depending on the mutual direction of the saturation magnetization and $k$-vector propagation \cite{dam61}. In case of DE mode bounded to upper interface (Fig.~\ref{f:sketch}(a)), the polar and longitudinal profiles of the dynamic magnetizations are \cite{dam61}
\begin{align}
\label{eq:mPde}
m_P^{\mathrm{(DE1)}}(z,t) &= \mPnull \exp(-\kswz z) \sin\omsw t
\\
\label{eq:mLde}
m_L^{\mathrm{(DE1)}}(z,t)&=\mPnull \epsilon \exp(-\kswz z) \cos\omsw t
\end{align}
where $\epsilon=\mLnull/\mPnull$ describes the ellipticity of the precessing magnetization, assumed to be constant over the whole FM layer thickness, $\kswz$ is the normal direction of the spin-wave wavevector. In the case of an homogeneous FMR mode, $\kswz=0$. Substituting the dynamic magnetization profiles (Eqs.(\ref{eq:mPde}-\ref{eq:mLde})) and the MOKE depth sensitivity functions (Eqs. (\ref{eq:dsP}-\ref{eq:dsL}) into the expression of MOKE effect (Eq.~\ref{eq:Phimode}), we get
\begin{equation}
\label{eq:MOKEde}
\Phi_{\s/\p}^{\mathrm{(DE1)}}(d,t)=
\mPnull P_{\s/\p}(0)\sqrt{1+\gamma_{\s/\p}^2\epsilon^2}
\,\,
\frac{1-\exp(-\kswz d - i \alpha)}%
{\kswz + i \alpha/d}
\,\,
\sin(\omsw t + \phi_{\s/\p})
\end{equation}
where $d$ is the thickness of the FM layer and $\alpha=4\pi N_z d/\lambda$. As the detected MOKE signal is a~mixture of both LMOKE and PMOKE, it results in   a~phase shift $\phi_{\s/\p}$ between the phase of the spin-wave mode and the detected MOKE signal, $\tan \phi_{\s/\p}=\gamma_{\s/\p}\epsilon$. 
Moreover, the term $\sqrt{1+\gamma_{\s/\p}^2\epsilon^2}$ originates from Pythagorean sum of PMOKE and LMOKE.

In the case of a~DE mode bounded to the lower FM interface (Fig.~\ref{f:sketch}(b)), the profiles of dynamic magnetizations are analogous to Eqs.~(\ref{eq:mPde}-\ref{eq:mLde})
\begin{align}
\label{eq:mPdebot}
m_P^{\mathrm{(DE2)}}(z,t) &= \mPnull \exp(-\kswz (d-z)) \sin\omsw\tau
\\
\label{eq:mLdebot}
m_L^{\mathrm{(DE2)}}(z,t)&=\mPnull \epsilon \exp(-\kswz (d-z)) \cos\omsw\tau
\end{align}
leading to the MOKE effect
\begin{equation}
\label{eq:MOKEdebot}
\Phi_{\s/\p}^{\mathrm{(DE2)}}(d,t)=
{\mPnull P_{\s/\p}(0)\sqrt{1+\gamma_{\s/\p}^2\epsilon^2}}
\,\,
\frac{\exp(- i \alpha)- \exp(-\kswz d)}
{\kswz - i \alpha/d}
\,\,
\sin(\omsw t + \phi_{\s/\p}).
\end{equation}

The last type of the spin-wave mode to be discussed here is the PSSW modes (Fig.~\ref{f:sketch}(c)), described approximately as a cosine function with its maxima pinned at the FM interfaces. Then, the amplitudes of dynamic magnetizations are
\begin{align}
\label{eq:mPpssw}
m_P^{\mathrm{(PSSW)}}(z,t)& = \mPnull \cos(m\pi z/d)  \sin\omsw\tau
\\
\label{eq:mLpssw}
m_L^{\mathrm{(PSSW)}}(z,t)&=\mPnull \epsilon \cos(m\pi z/d) \cos\omsw\tau
\end{align}
where integer $m$ denotes the mode number of a given PSSW mode. Substituting those magnetization profiles (Eqs.~(\ref{eq:mPpssw}-\ref{eq:mLpssw})) to Eq.~(\ref{eq:Phimode}), leads to the MOKE effect
\begin{equation}
\label{eq:MOKEpssw}
\Phi_{\s/\p}^{\mathrm{(PSSW)}}(d,t)=
{ \mPnull P_{\s/\p}(0)\sqrt{1+\gamma_{\s/\p}^2\epsilon^2}}
\,\,
\frac{%
i\alpha d \left[1-\exp(- i \alpha) (-1)^m\right]
}{m^2\pi^2 - \alpha^2}
\,\,
\sin(\omsw t + \phi_{\s/\p}).
\end{equation}

Please recall, that the analytical expression of the BLS intensity $I_{\s/\p}^{(\oml\pm\omsw)}$ of different modes can be  easily obtained from the complex Kerr angle $\Phi_{\s/\p}$  using Eq.~(\ref{eq:MOKEBLS}), $I_{\s/\p}^{(\oml\pm\omsw)} = I_0 |\Phi_{\s/\p} r_{\s\s/\p\p}|^2$, where $\Phi_{\s/\p}$ is given by Eqs.~(\ref{eq:MOKEde}), (\ref{eq:MOKEdebot}), (\ref{eq:MOKEpssw}), and where the sign of LMOKE must be reversed, e.g.\ by reversing the sign of $\gamma_{\s/\p}$ \emph{or} $m_L$.

\section{MOKE and BLS signal from spin-waves in a N\lowercase{i} film}
\label{Sec:V}

\begin{figure}
\begin{center}
\includegraphics[width=0.6\textwidth]{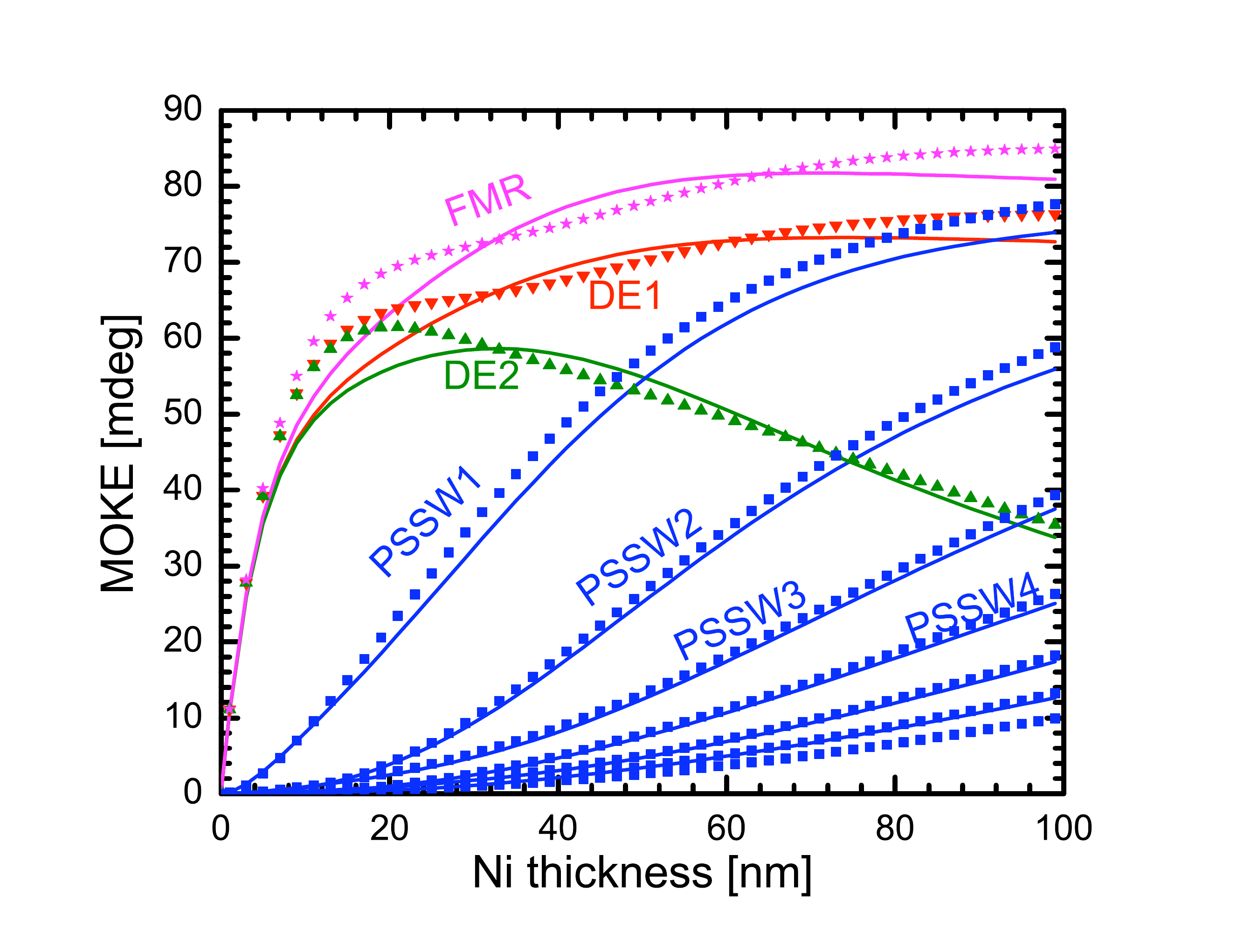}
\end{center}
\caption{%
\label{f:analyt}%
(color) MOKE signals $|\Phi|$ originating from the different types of spin waves as a function of the Ni thickness, assuming the spin-wave ellipticity $\epsilon=2$. (\textcolor{red}{$\blacktriangledown$}, \textcolor{green}{$\blacktriangle$}) denotes for the DE modes bounded to upper and lower FM interface, respectively, with $\kswz=10^{7}$\,m$^{-1}$. (\textcolor{magenta}{$\star$}) denotes for the FMR mode (i.e.\ as the DE modes with $\kswz=0$). (\textcolor{blue}{$\blacksquare$}) denotes for the PSSW modes. Incidence angle is 25$^\circ$, light wavelength is 810\,nm. For values of optical and magneto-optical parameters, see text. Symbols are the optically exact $4\times4$ matrix formalism, solid lines the analytical expressions (Eqs.~(\ref{eq:MOKEde}, \ref{eq:MOKEdebot}, \ref{eq:MOKEpssw})).
}
\end{figure}

Figure~\ref{f:analyt} shows the MOKE effect from  various spin-wave modes as a function of the Ni film thickness in air/Cu(2\,nm)/\baf Ni($d$)/\baf Cu(5\,nm)/\baf Si structure. The investigated modes are the DE mode, bounded to both the upper (DE1) and the lower (DE2) interface, expressed for $\kswz=10^{7}$m$^{-1}$. Furthermore, the investigated modes are the FMR mode and several PSSW modes with different mode numbers. Within those calculations, for demonstration purposes, we keep the spin-wave ellipticity $\epsilon$ constant, although obviously the spin-wave ellipticity is different for different spin-wave modes.

(i)
First, the calculations demonstrate a nice agreement between the exact optical calculations based on $4\times4$ matrix formalism and analytical formulae (Eqs.~(\ref{eq:MOKEde}), (\ref{eq:MOKEdebot}), (\ref{eq:MOKEpssw})). The largest disagreement is for Ni thicknesses in the range of about 3 -- 30\,nm, as for this range the analytical expressions of $L_{\s/\p}(z)$, $P_{\s/\p}(z)$, as given by Eqs.~(\ref{eq:dsP})-(\ref{eq:dsL}), are not exact, as already discussed in Sec.\ref{Sec:III}.

(ii)  As the Ni thickness is increasing, the MOKE signals increase and then saturate. In Fig.~\ref{f:analyt}, the saturation is clearly visible only for the FMR, DE1 and PSSW1 modes. The saturation roughly appears when the depth of the unique sign of the dynamic magnetization corresponds to the length $2\LambdaMOKE$.
In the case of the FMR or DE1 mode, the MOKE signal saturates roughly at $2\LambdaMOKE\approx d$.
Within our Ni example, $2\LambdaMOKE\approx 30$\,nm and hence the MOKE signal saturates roughly at $d=30$\,nm for the FMR/DE1 mode, in agreement with Fig.~\ref{f:analyt}.
For PSSW modes, a unique sign of the dynamic magnetization holds for $1/4$ of the perpendicular spin wavelength, which is $\lambdaperp\approx 2d/m$ ($m$ is integer denoting the PSSW mode order). Consequently, the PSSW modes saturate when $(1/4)\lambdaperp= d/(2m)\approx2\LambdaMOKE$. Hence the estimated PSSW mode saturation appears at $d=4m\LambdaMOKE$, which corresponds to $60$, $120$, $180$\,nm, etc.\ for PSSW modes with $m=1$, 2, 3, etc., respectively. This is also in qualitative agreement with the calculations shown in Fig.~\ref{f:analyt}.

(iii) For a large thickness of the FM layer ($d\rightarrow\infty$), the MOKE signal from DE1/FMR and PSSW modes saturate to the same value, being $\Phi_{\s/\p}(d,t)=\mPnull P_{\s/\p}(0)\sqrt{1+\gamma_{\s/\p}^2\epsilon^2} (\lambda/4\pi i N_z) \sin(\omsw t + \phi_{\s/\p})$, as follow from Eqs.~(\ref{eq:MOKEde}), (\ref{eq:MOKEpssw}). The underlying physics is that the upper part of the  FM layer, which is probed by light, has nearly constant amplitude of the dynamic magnetization. Hence, obviously, in case of PSSW modes, such saturation appears for larger thicknesses of the FM layer, as compared to DE1/FMR modes.

(iv) For smaller Ni thicknesses, MOKE signals for PSSW modes are strongly reduced with increasing number of the PSSW modes. The reason is analogous to the discussion in point (ii): when the depth profile of the PSSW mode oscillates  on a short distance as compared to the probing depth of MOKE, $\LambdaMOKE$, then MOKE contributions from various depths of the Ni film cancel each other.
Quantitatively, when 1/4 of the perpendicular spin-wavelength, $(1/4)\lambdaperp=d/(2m)$, is shorter than half of the MOKE probing length, $\LambdaMOKE/2$, then the provided MOKE signal is significantly reduced. Substituting $\LambdaMOKE=15$\,nm, the reduction of MOKE signal is for Ni thicknesses below 15, 30, 45\,nm for PSSW1, PSSW2, PSSW3, respectively, in agreement with Fig.~\ref{f:analyt}.

\begin{figure}
\begin{center}
\includegraphics[width=0.8\textwidth]{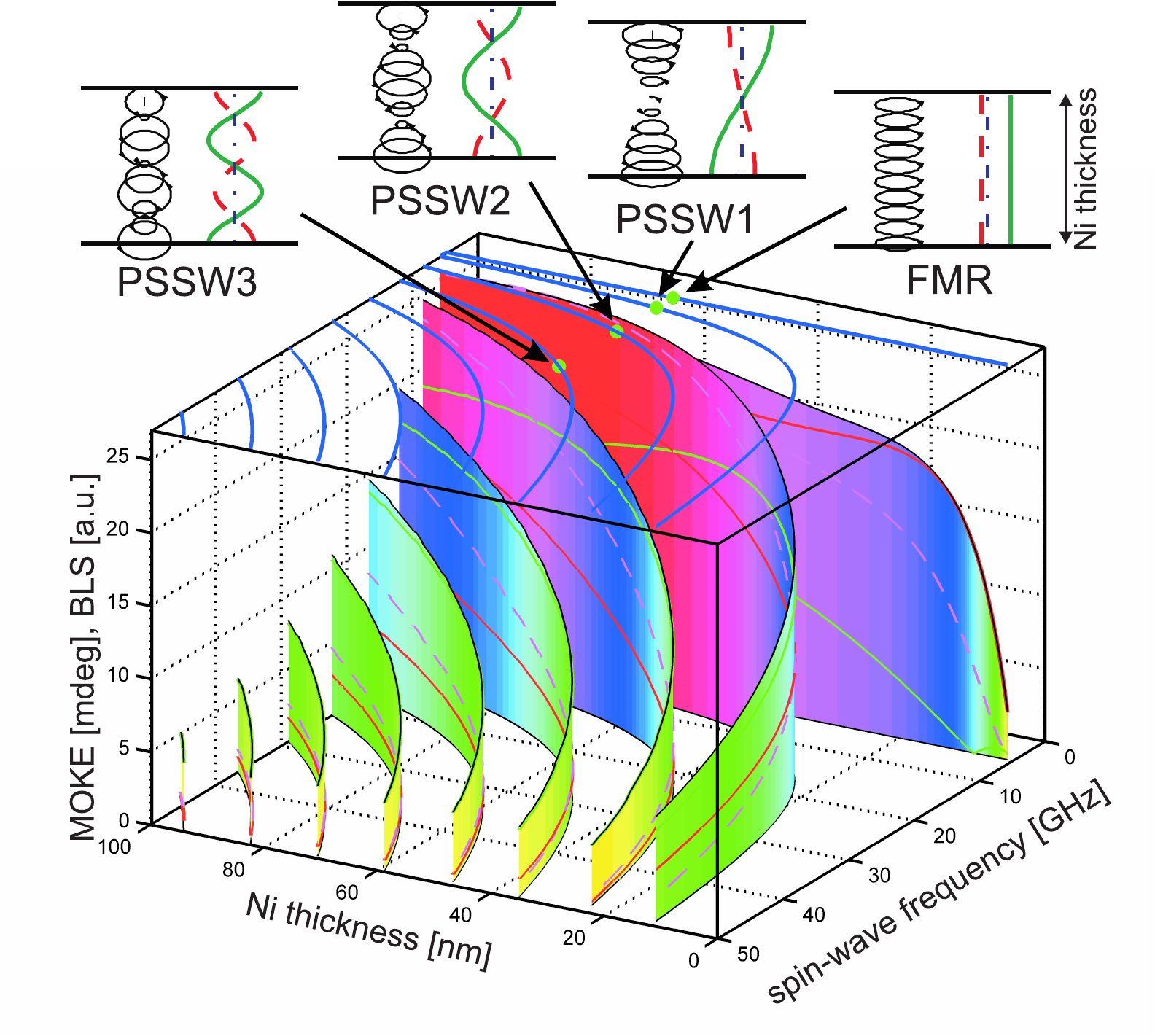}
\end{center}
\caption{%
\label{f:disp}
(color) Calculated MOKE signals (height of the surface plot: $|\Phi|$, solid red line: Kerr rotation $\theta=\mathrm{Re}(\Phi)$, solid green line: Kerr ellipticity $\epsilon=\mathrm{Im}(\Phi)$) and scaled BLS signal (dashed magenta line) as a function of the Ni thickness and the related spin-wave frequencies. For values of optical and magnetic parameters, see text. The top plane of the 3D graph presents the spin-wave dispersion of the Ni film (solid lines). The profiles of the spin-wave amplitudes through a~60-nm-thick Ni film are sketched above the 3D graph for four examples of the spin-wave modes, one FMR and three PSSW modes. Within those profiles, the left part is showing the trajectory of the end-point of the magnetization vector. The right part shows the amplitudes of the dynamic in-plane (solid line) and normal (dashed line) magnetizations.}
\end{figure}

Figure \ref{f:disp} demonstrates MOKE and BLS signals originating from the spin waves inside the Ni film, where the spin-wave profiles and frequencies are calculated using phenomenological models \cite{hil90,buch03}. The used magnetic properties of Ni are: saturation magnetization $\mu_0M_S=659$\,mT (i.e.\ $M_S=520$\,kA/m), exchange constant $A=4.7$\,pJ/m (i.e.\ exchange stiffness $D=2.46$\,meV\,nm$^2$), Land\'e $g$-factor $g=2.1$, out-of-plane anisotropy $K_z=0$\,kJ/m$^3$ \cite{lenunpub}. The in-plane spin-wave wavevector is $q_\|=0$, and the external magnetic field $\mu_0H=50$\,mT is applied parallel to the plane of incidence. The optical parameters of the Ni film as well as its optical surrounding are described above in Sec.~\ref{Sec:III}.
The dependence of the MOKE signals on the Ni thickness and the related spin-wave frequencies are presented in Fig.~\ref{f:disp} for the FMR mode as well as for several PSSW modes. The MOKE signals are expressed as $|\Phi|$, Kerr rotation $\theta=\mathrm{Re}(\Phi)$ and Kerr ellipticity $\epsilon=\mathrm{Im}(\Phi)$. The amplitude of the in-plane dynamic magnetization of the FMR mode is chosen to be one, $m_L^{\mathrm{(FMR)}}=1$. The amplitudes of the dynamic magnetizations of the other spin wave modes are normalized in a way that their energies per unit area are equal  \cite{coch88,buch07}.
%-- for example in case of thermal spin-wave this energy density is proportional to $kT$ \cite{coch88,buch07}.

All expressions of the MOKE signals ($|\Phi|$, $\theta$, $\epsilon$) provide very similar dependencies.
As the external field $H$ was applied parallel to the plane of incidence, the longitudinal contribution to the MOKE effect is zero, as there is no dynamical magnetization in the longitudinal direction. However, as discussed in Sec.~\ref{Sec:III}, the longitudinal depth sensitivity function is here about 12-times smaller than the polar one. As the normal spin-wave amplitudes are about 1--2-times smaller than the in-plane amplitudes, the polar contribution would be dominant even in the case of $H$ perpendicular to the plane of  incidence, where the LMOKE contributes to the outgoing MOKE signal.

The MOKE signals in Fig.~\ref{f:disp} are compared to the normalized BLS intensity (dashed magenta line). As the BLS signal is basically square of the MOKE signal (Eq.~(\ref{eq:MOKEBLS})), the behaviour of  scaled BLS  and MOKE signals are very similar, while the BLS signal is reduced more significantly for smaller Ni thicknesses or for higher orders of the PSSW modes.

\section{Conclusion}

In conclusion, we have analytically expressed the MOKE and BLS signals originating from prototypical spin-wave modes, namely the Damon--Eshbach (DE) and Perpendicular Standing Spin Wave (PSSW) modes. The calculations are based on the additivity of the MOKE effect, on analytical expression of the depth sensitivity function of the MOKE signal, as well as on a~straightforward relation between the MOKE and BLS signals. As a showcase, we have expressed the MOKE and BLS signal in a~Ni film. It is shown that analytical calculations describe well the physical behavior, as follows from the comparison with exact magneto-optical calculations. Furthermore, we have demonstrated that the dependence on the FM-layer thickness  of both the MOKE and BLS signals is very similar. Namely, with increasing FM layer thickness, the MOKE and BLS signals saturate, where for PSSW modes with higher mode number, the saturation is provided for larger FM layer thicknesses. Furthermore, the MOKE and BLS signals reduce significantly as the PSSW mode number is increasing.

\section{Acknowledgment}

The financial support through Grant Academy of the Academy of Sciences of the Czech Republic (KAN400100653), Grant of Ministry of Education, Youth and Sports of the Czech Republic (MSM6198910016) and the EU -- Regional Materials Science and Technology Centre (CZ.1.05/2.1.00/01.0040) are gratefully acknowledged.

%\bibliography{depth}
%%\bibliographystyle{prsty}
%\bibliographystyle{vancouver-jaro}

%%%%%%%%%%%%%%%%%%%%%%%%%%%%%%%%%%%%

%%%%%%%%%%%%%%%%%%%%%%%%%%%%%%%%%%%%
\clearpage

\end{document}